\documentclass[12pt]{iopart}
\usepackage{graphicx}

\usepackage{mathrsfs}
\usepackage{amsthm}
\usepackage{amssymb}
\usepackage{hyperref}
\usepackage{xcolor}
\usepackage{cite}
\usepackage{dcolumn}
\usepackage{bm}

\usepackage{iopams}  

\newcommand{\x}{{\bf r}}
\newcommand{\K}{{\bf k}}
\newcommand{\q}{{\bf q}}

\begin{document}

\title{Quantum droplets in a dipolar Bose gas at a dimensional crossover}


\author{Pawe{\l} Zin$^{1,2}$, Maciej Pylak$^{1,3}$, Tomasz Wasak$^4$, Krzysztof Jachymski$^{2,5}$ and Zbigniew Idziaszek$^2$   }
\address{$^1$ National Centre for Nuclear Research, ul. Pasteura 7, PL-02-093 Warsaw, Poland}
\address{$^2$ Faculty of Physics, University of Warsaw, ul. Pasteura 5, PL-02-093 Warsaw, Poland}
\address{$^3$ Institute of Physics, Polish Academy of Sciences, Aleja Lotnik\'ow 32/46, PL-02-668 Warsaw, Poland}
\address{$^4$ Max Planck Institute for the Physics of Complex Systems, N\"othnitzer~Str.~38, D-01187 Dresden, Germany }
\address{$^5$ Institute for Quantum Control (PGI-8), Forschungszentrum J\"ulich, D-52425 J\"{u}lich, Germany}

\vspace{10pt}

\begin{abstract}
We study the beyond-mean-field corrections to the energy of a dipolar Bose gas confined to two dimensions by a box potential with dipoles oriented in plane such that their interaction is anisotropic in the two unconfined dimensions. At a critical strength of the dipolar interaction the system becomes unstable on the mean field level. We find that the ground state of the gas is strongly influenced by the corrections, leading to formation of a self-bound droplet, in analogy to the free space case. Properties of the droplet state can be found by minimizing the extended Gross-Pitaevskii energy functional. In the limit of strong confinement we show analytically that the correction can be interpreted as an effective three-body repulsion which stabilizes the gas at finite density.
\end{abstract}

\submitto{\jpb}
%
%
%
%
%

\section{Introduction}
Ultracold dipolar gases have been attracting great attention in recent years. In these systems, long-range anisotropic dipole-dipole interaction gives rise to a plethora of
novel effects which makes them useful for various applications in quantum simulation~\cite{baranov2012condensed}. Different aspects of such systems can be studied using a variety of experimental platforms including magnetic atoms~\cite{Goral2000,Lahaye:2009}, polar molecules~\cite{bohn2017cold} and Rydberg atoms~\cite{saffman2010quantum},
allowing to access different regimes of interaction strength, geometry, particle number and quantum statistics. On the mean field level, a dilute gas of dipolar bosons
can become unstable towards a collapse caused by partially attractive nature of the interaction~\cite{Lahaye:2009}. This can be seen also in the excitation spectrum as the
Bogoliubov mode frequencies become imaginary. Depending on the external confinement, the instability can be tuned to occur at different values of the dipolar interaction
strength and change its character.

The development of experimental techniques allowing to produce Bose condensed clouds of highly magnetic lanthanide atoms such as erbium and
dysprosium~\cite{lu2011strongly,aikawa2012bose} has led to rapid progress in this field. Most notably, the experiments highlighted the role of beyond-mean-field effects
for the dynamics of the gas with the unexpected discovery of the droplet state~\cite{Kadau2016,Barbut2016,Chomaz2016,Schmitt2016,Wenzel2017}. It turns out that close to
the instability, the mean field contribution to the energy of the gas vanishes and the beyond-mean-field corrections, which typically have higher power-law dependence on
the density, become important. The positive correction to the chemical potential can be interpreted as a source of effective repulsion in the gas. This
results in formation of a long-lived finite size droplet with liquid properties. This kind of quantum droplet was originally suggested to occur in Bose-Bose
mixtures~\cite{Petrov2015,Petrov2016}, and was later observed as well ~\cite{Semeghini2018,Cabrera2018}.

The discovery of quantum droplets resulted in renewed theoretical interest in calculations of the beyond-mean-field corrections, pioneered already many years
ago~\cite{Lee1957,Beliaev1958,Schick1971,hugenholtz1959ground}. For dilute Bose gas with short-range interactions the first results have been provided by Lee, Huang and
Yang (LHY)~\cite{Lee1957}. For the case of dipolar interactions in free space, the correction turns out to have the same dependence on the density of the gas, but its
magnitude is enhanced~\cite{Pelster2012}. The presence of external confining potential enriches the problem by introducing a new lengthscale and allowing for the
effective reduction of the system dimensionality, which modifies the functional form of the beyond mean field terms~\cite{Popov,Mora2009}.  For anisotropic interactions,
the relative orientation between the dipoles and trap geometry can be exploited to qualitatively change the excitation spectrum, developing a roton
mode~\cite{Santos2003,Fischer2006,Boudjemaa2013,Chomaz2018,Petter2018,Kora2019}.
The possibility to modify the properties of the roton mode allows to explore novel phases of matter.
As demonstrated recently, by carefully tuning the parameters it was possible to bring the ground state of the system from a
single droplet to an array of phase-coherent droplets~\cite{Bombin2017,Cinti2017,Tanzi2018,Roccuzzo2019,Bottcher2019,Zhang2019}, featuring broken translational
symmetry along with superfluid order and, thereby, having the properties of a supersolid~\cite{Boninsegni2012,li2017stripe,leonard2017supersolid}. Quantum droplets
were also predicted to occur in dipolar bosonic mixtures \cite{Smith2021,Bisset2021} and in Rabi-coupled Bose mixtures \cite{Cappellaro2017}.

Theoretical studies of the droplet physics were so far largely restricted to solving an extended Gross-Pitaevskii (GP) equation with an additional term accounting for the
LHY correction~\cite{Barbut2016,Wachtler2016,bisset2016ground} taken from free-space three-dimensional calculation~\cite{Pelster2012}. The validity of this approach
relies on the local density approximation (LDA), whereas external confinement is known to modify the beyond-mean-field
corrections~\cite{Edler2017,zin2018droplets,buechler2018crossover,jachymski2018nonuniversal}. Predictions of the extended GP equation have so far been rather successful
in interpreting the experimental results, and have been to some extent supported by Monte Carlo calculations~\cite{Macia2016droplets,Cinti2017superfluid,Bottcher2019a}.
The aim of this paper is to rigorously study the LHY term in a confined dipolar system, and to provide the form of the LHY correction for an effectively two-dimensional
system, as well as to check the validity of LDA.  The calculation can be performed mostly analytically provided that the confining potential is assumed to be of a box
type with periodic boundary conditions.

This work is structured as follows. In Section~\ref{sec:description}, we introduce the system. 
In Section~\ref{sec:LHY} we perform the calculation of the LHY correction for the uniform system.
In Section~\ref{sec:droplet}, we show that the calculated correction can give rise to formation of droplets. Conclusions are drawn in Section~\ref{sec:conclusion}. 
Technical details of the calculations  are given in the Appendices.

\section{Description of the system}\label{sec:description}

\begin{figure}[t]
  \centering
  \includegraphics[width=.45\textwidth]{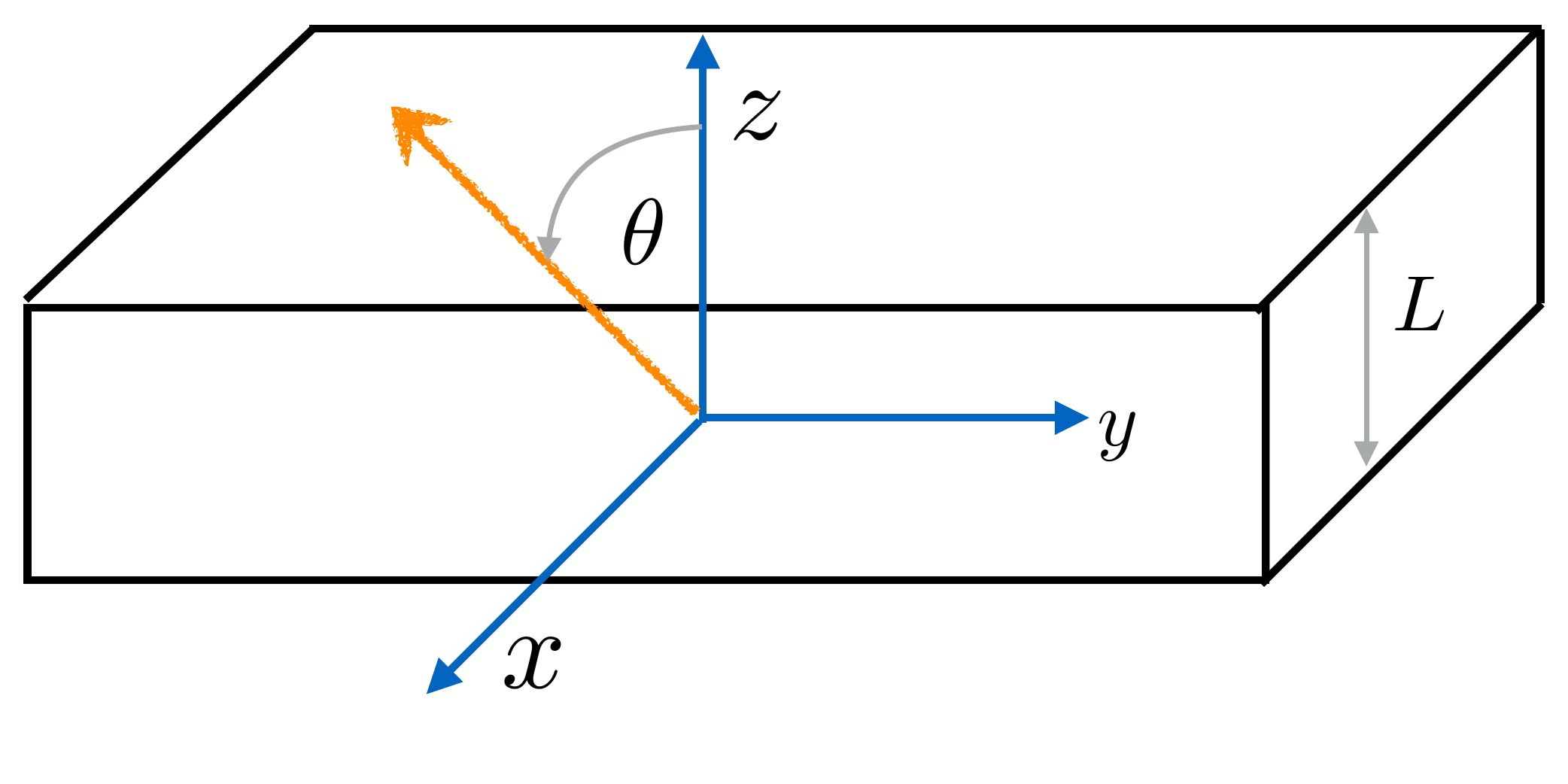}
  \caption{Graphical illustration of the system geometry. The dipoles (thick orange arrow) lie in the $x$--$z$ plane and are tilted by the angle $\theta$ with respect to the $z$-axis.
    The gas is confined in the $z$-direction in a box of length $L$.
  }\label{fig:coord}
\end{figure}

The system under study is an ultradilute gas of polarized bosonic dipoles confined in a highly anisotropic trap. For simplicity, we consider the system being placed between two infinite planes separated by distance $L$ from each other. 
We choose the coordinate system such that $z$ is the direction perpendicular to the planes.
The dipoles polarization direction is taken to be tilted by angle $\theta$ from the $z$ direction
(see Fig.~\ref{fig:coord}). 
We assume that the system forms a Bose-Einstein condensate, and study its properties using the standard Bogoliubov method. Than the energy of the system reads
\begin{eqnarray} \nonumber
  E[\psi]\!=\!&&\int\!\! d\x_\perp \int_{-L/2}^{L/2}  d z \, \frac{\hbar^2}{2m} |\nabla \psi|^2
\\ \label{energia}
&&
+\! \frac{1}{2}\int \!\!d\x_\perp d\x_\perp'\,  \int_{-L/2}^{L/2}\!\!\! dz dz' \,   v(\x-\x')
|\psi(\x)|^2 |\psi(\x')|^2  \!  + E_{LHY}[\psi]
\end{eqnarray} 
where $v(\x-\x')$ denotes the interaction potential and $E_{LHY}$ the LHY energy term.

We additionally assume that the gas has a constant density in the $z$ direction.  This naturally restricts our considerations to the systems where the changes of the density in the
$x$-$y$ plane takes place on distances much larger than $L$. Below, when we derive the profile of the droplet, we exploit the variational approach for the wave function in the
$x$-$y$ plane, but still the $z$ component is position independent~\cite{martin2017vortices}.  With such approximation Eq.~(\ref{energia}) takes the form
\begin{eqnarray} \nonumber
  E[\psi_\perp]\!=&&\!\int\!\! d\x_\perp \, \frac{\hbar^2}{2m} |\nabla_\perp \psi_\perp|^2 
\\ \label{energia2}
&&
+\! \frac{1}{2}\int \!\!d\x_\perp d\x_\perp'\,   v_{2d}(\x_\perp-\x_\perp')
|\psi_\perp(\x_\perp)|^2 |\psi_\perp(\x_\perp')|^2  \! + E_{LHY}[\psi_\perp] 
\end{eqnarray}
where $\x_\perp = x {\bf e}_x + y {\bf e}_y$ we denote the two-dimensional position vector,
$\psi_\perp = \sqrt{L} \psi$ and
\begin{equation}\label{eq:v2d}
   v_{2d}(\x_\perp-\x_\perp') = \int_{-L/2}^{L/2}\!\!\! dz dz' \, v(\x-\x').
\end{equation}

\section{Beyond mean field energy correction}\label{sec:LHY}

After describing the system, we proceed to the calculation of the energy $E_{LHY}$.
To make the system analytically tractable, we use the local density approximation, i.e., we calculate the LHY energy density of a uniform
system $\frac{\epsilon_0}{L^2} e_{LHY}^{2d}$ and then approximate
\begin{equation}\label{LDA}
E_{LHY}[\psi_\perp] \simeq \int\!\! d\x_\perp \,   
 \frac{\epsilon_0}{L^2} e_{LHY}^{2d}\!\left[\frac{2}{\pi} aL |\psi_\perp(\x_\perp)|^2 \right].
\end{equation}
Here, we introduced an energy scale $\epsilon_0 \equiv \frac{\hbar^2}{2m} \left( \frac{2\pi}{L} \right)^2$. We separate out the factors
$\frac{\epsilon_0}{L^2}$ and $\frac{2}{\pi} aL$ for convenience and brevity of further formulas.

In order to evaluate the energy $ \frac{\epsilon_0}{L^2} e_{LHY}^{2d}$, we work with periodic boundary conditions (PBC) along $z$, which is an approximation that significantly
simplifies the calculations.  Although convenient for analytical considerations, mathematically, they introduce spurious interaction between the Fourier copies of the
system in adjacent cells of the periodic system since the dipole-dipole interaction has a long tail. However, it was shown in Ref.~\cite{Ronen2006} that this spurious
interaction modifies the mean field energy by less than one percent as compared to the case without PBC. Since in the 3D limit, the local density approximation was
successfully used with the LHY energy in the form of the generalized Gross-Pitaevskii equation, we expect that the LHY energy, being more local in nature, will also be
little affected by the spurious interaction.


\subsection{Interaction potential}\label{sec:potential}

To proceed, we first focus on the interaction potential $v(\K)$. 
The bare potential consists of two parts. The first is the anisotropic dipole-dipole
interaction, and the second is a strong, short-range potential which we assume to be isotropic and dominating over the dipolar part at small distances. However, as long
as only low energy and large distances are considered, the pseudopotential of the following form can be used~\cite{Yi2000,Oldziejewski2016}:
\begin{equation} \label{v}
   v(\mathbf{r})=g\bigg[\delta(\mathbf{r})+\frac{3\epsilon_{dd}}{4\pi r^3}\left(1-3 ({\bf e}_\x \cdot {\bf e})^2 \right)\bigg].
\end{equation} 
Here the interaction strength $g$ is determined by the scattering length of the total potential $a$ and atomic mass $m$ by $g=\frac{4\pi\hbar^2 a}{m}$ and $\epsilon_{dd}$
parametrizes the relative strength of the dipolar part of the interaction with respect to the short-range one, which can be expressed as a ratio $a_{\rm dd}/a$ with $a_{\rm dd}$
being a characteristic dipolar length.  In addition, ${\bf e}_\x = \x/|\x|$ and ${\bf e}$ denotes the polarization direction.  We emphasize that $a$ is the scattering length of the
total potential, i.e., the sum of the short-range and the dipolar potentials, and it may differ from the scattering length of the short-range potential
significantly~\cite{Ronen2006druga,Bortolotti2006}.

Since we calculate LHY energy correction in a system with periodic boundary conditions, we 
shall make use of Fourier transform of the $v(\x)$ potential given by 
\begin{equation}\label{gvk}
   v(\K) = \int_{-L/2}^{L/2}\!\!dz  \int \!\!d\x_\perp \, e^{-i \K \x} \frac{v(\x)}{g}
\end{equation}
where $\x_\perp = x {\bf e}_x + y {\bf e}_y$. Due to periodic boundary conditions
$k_z$ is quantized, i.e., $k_z = \frac{2\pi}{L} q_z$ where $q_z$ is an integer.
To simplify further formulas, we use in the above $1/g$ prefactor.
We obtain the following analytical form of the potential
\begin{equation} \label{potdd} 
\fl
 v(\q) = 1  
+  \epsilon_{dd} \bigg\{
3   \frac{ q_x^2  \sin^2 \theta + q_x q_z \sin 2 \theta -  q_\perp^2 \cos^2 \theta  }{q^2} \,
    \big[ 1 - (-1)^{q_z} e^{-\pi q_\perp} \big] +3 \cos^2 \theta  -1 \bigg\},
\end{equation}
where  $\q = \K L /2\pi$, $q_\perp = \sqrt{q_x^2+q_y^2}$, and we assume, without the loss of generality, the dipoles to be oriented within the $x-z$ plane.  Here, $\theta$ is the angle
between the polarization direction and the $z$-axis, see Fig.~\ref{fig:coord}.  Since the dipole-dipole potential scales as $1/r^3$ with the distance, some caution needs
to be taken at the two limiting cases: $r \to 0$ and $r \to \infty$. We discuss this issue in \ref{app0} for the sake of completeness. 
The above formula has also been calculated in \cite{Ronen2006,Baillie2015}. However, there $L$ had a meaning
of a cutoff in the position space. Therefore, $k_z$ present in these works are not quantized.

\subsection{Critical point}\label{sec:critical}


We now proceed to the analysis of the critical point of the uniform system defined above.
We define it as the instability point of the spatially uniform solution within the
mean field approximation. We assume that initially the gas is homogeneous and search for the instability with increasing $\epsilon_{dd}$. This
instability can be found by analysing the spectrum of the Bogoliubov quasiparticles
$\varepsilon(\q) = \sqrt{ q^2[q^2 + 2 \xi v(\q)] } = \sqrt{q^2 f(\q)}$ where $f(\q) = q^2 + 2 \xi v(\q) \geq 0$. To simplify the notation we introduce $\xi = gn/\epsilon_0$ as the dimensionless parameter measuring the strength of the interactions. 
In the stable phase, all the quasiparticle energies should be real.  However, at the critical point,
the function $f(\q) = q^2 + 2 \xi v(\q) \geq 0$, while for at least a single value of $\q = \q_c$, we have $f(\q_c) = q_c^2 + 2 \xi v(\q_c) = 0$.  It is straightforward
to find that for $|\delta \theta | \ll 1 $ where $\delta \theta = \theta - \pi/2$ the above shall hold for $q_z=0$. In other words, the instability occurs in the plane,
where dipolar attraction is the strongest. In such a case, according to Eq.~(\ref{potdd}) we have
\begin{eqnarray*}
f(\q) &=&
q^2 
+ 2 \xi \left( 1 + \epsilon_{dd,crit} (  3 \cos^2 \theta  -1   ) \right)
\\
&+&
   6\xi  \epsilon_{dd,crit} (\sin^2 \phi  \sin^2 \theta - \cos^2 \theta)  \,
    \big[ 1 -  e^{-\pi q_\perp} \big], 
\end{eqnarray*}
where $\sin \phi = q_x/q_\perp$. The minimal value is achieved for $\phi =0$. At the critical point $\q=\q_c$ we have $f(\q) =0$ and $\nabla_\q f(\q) = 0$, which leads to
\begin{eqnarray*}
q_c^2 +2 \xi (1 - \epsilon_{dd,crit})
+
 6 \xi  \epsilon_{dd,crit}   \sin^2 \delta \theta   \, e^{-\pi q_c}    = 0 
\end{eqnarray*}
and
\begin{eqnarray*}
q_c e^{\pi q_c} =  3  \pi \xi  \epsilon_{dd,crit}    \sin^2 \delta \theta     \, .
\end{eqnarray*}
The above can be rewritten as
\begin{equation}\label{epcrit}
    \epsilon_{dd,crit} - 1 = \frac{1}{2\xi} \left(  q_c^2  + \frac{2}{\pi} q_c \right)  \, ,
\end{equation}
where $q_c$ is a solution of
\begin{equation}\label{qceq}
q_c e^{\pi q_c} =  3  \pi \xi  \sin^2 \delta \theta
\left(      1 + \frac{1}{2\xi} \left(  q_c^2  + \frac{2}{\pi} q_c \right)   \right).
\end{equation}
As we can see, in general $\epsilon_{dd,crit}$ depends on $\delta \theta$ and $\xi$.
However, there are special cases that we want to address.
Firstly, for $\delta \theta =0$ the solution simplifies to $q_c = 0$ and $\epsilon_{dd,crit} = 1$ which does not depend on the value of $\xi$.
There is also another interesting limit where $\delta \theta \ll 1$ and simultaneously $ \delta \theta^2 \xi \ll 1$, where we find that 
\begin{equation}\label{epp_crit}
\epsilon_{dd,crit} \simeq 1 + 3 \delta \theta^2 
\end{equation}
which again shows lack of  $\xi$ dependence.


\subsection{The beyond-mean-field term calculation}\label{sec:lhy}

We proceed to the calculation of the analogue of the Lee-Huang-Yang term for the trapped gas interacting with the potential given by Eq.~(\ref{v}) in the case of uniform system.
We express Lee-Huang-Yang $3d$ energy density of the system as $\frac{\epsilon_0}{L^3} e_{LHY}^{2d}(\epsilon_{dd},\theta,\xi)$, and~\cite{HP1959}
\begin{equation}
  \label{lhydef}
  -\frac{2 e_{LHY}^{2d}}{\xi^2} \!=\!  \sum_{q_z}\!\! \int\!\! d\q_\perp   \frac{v^2(\q)}{ \varepsilon(\q) + q^2 + 
  \xi v(\q)} \!-\!  \int\!\! d\q \frac{ v_{3\mathrm{d}}^2(\q)}{2q^2}\, ,
\end{equation}
where $\q_\perp = q_x {\bf e}_x + q_y {\bf e}_y$, $v(\q)$ is given by Eq.~(\ref{v}), $\varepsilon(\q) = \sqrt{ q^2[q^2 + 2 \xi v(\q)] }$ and 
$v_{3d}(\q) = \int \mbox{d} \x \, \exp \left( - i \frac{2\pi}{L} \q \x \right) v(\x)/g = 1 + \epsilon_{dd} \left(3 \frac{(\q \cdot {\bf e})^2}{q^2} - 1 \right)$  is the three-dimensional Fourier transform of the potential $v(\x)$. Here ${\bf e} = \sin \theta {\bf e}_x +  \cos \theta {\bf e}_z$ is the direction of the dipoles' polarization (see Fig.~\ref{fig:coord}).
The second term in Eq.~(\ref{lhydef}) results from the standard high momenta renormalization procedure which we describe in \ref{appR}.  
For $\xi \gg 1$, the atoms occupy many excitation levels in the confined direction and the system should behave as three-dimensional. Indeed, we find that for $\xi\gg1$, the
beyond-mean-field energy $e_{LHY}^{2d}$ recovers the 3D result~\cite{Pelster2012}
\begin{eqnarray}
\label{lhy3d}
e_{LHY}^{2{d}}(\xi) \stackrel{\xi\to\infty}{\longrightarrow}e_{LHY}^{3{d}}= \xi^{5/2} \frac{8\pi \sqrt{6}}{5}\, .
\end{eqnarray}

\subsubsection{Critical point calculation}

\begin{figure}[htb]
  \centering
\includegraphics[width=0.49\textwidth]{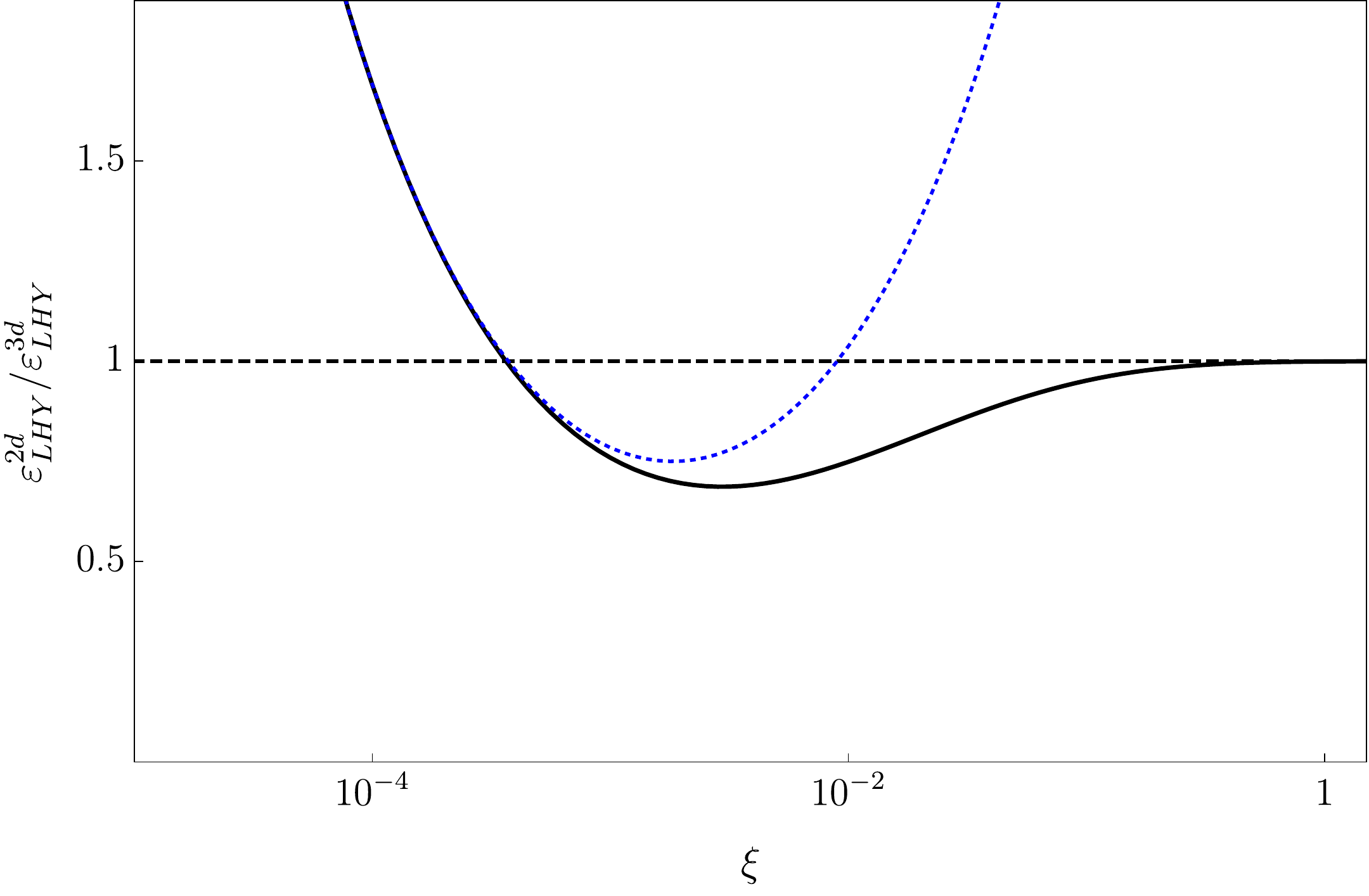}
\includegraphics[width=0.49\textwidth]{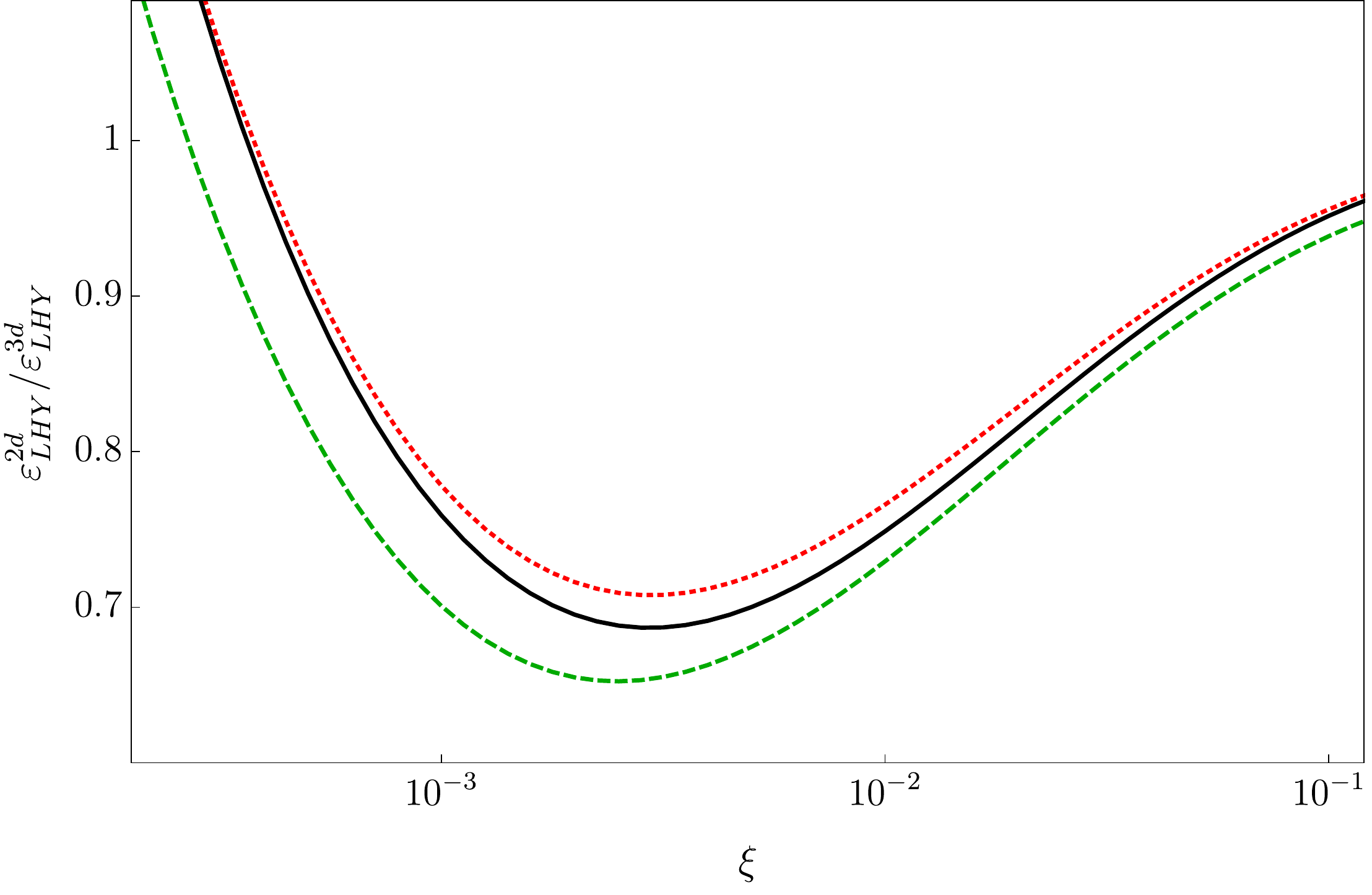}
  \caption{Left panel: the LHY energy $e_{LHY}^{2{d}}$ as a function of $\xi$ at the critical point. The energy (solid line) is compared to the 3D limit from Eq.~(\ref{lhy3d}) (dotted line). 
    The dashed blue line shows the result of the approximation from Eq.~(\ref{lhy2d}.
Right panel: a magnified view of the comparison of the LHY energies calculated for $\delta \theta = 0$, $\epsilon_{dd} = 1$ (black), $\delta \theta = 0$,   $\epsilon_{dd} =1 - 0.01$ (green dashed) and  $\delta \theta = 0.1$, $\epsilon_{dd} = 1$ (red dotted).}
  \label{fig2}
\end{figure}

We now focus on the properties of $e_{LHY}^{2d}(\xi)$ for $\delta \theta =0 $ and $\epsilon_{dd}=1$, which as described in Section~\ref{sec:critical}, is the critical
point. Fig.~\ref{fig2} shows the numerically calculated results for the corrections described by Eq.~(\ref{lhydef}) as a function of~$\xi$. The beyond-mean-field term
approaches the 3D limit with increasing $\xi$. As can be seen from the figure, the value of $e_{LHY}^{2{d}}$ is already close to the limiting case of $e^{3{d}}_{LHY}$ for
$\xi \gtrsim 0.1$. This means that a rather strongly confined gas can still be reasonably well described using the free space results with the trap incorporated in the
LDA fashion.

The $\xi\ll 1$ regime describes the quasi-2D limit in which the collisions are 3D in character but the interaction energy is too low to populate the excited states in the
confined direction. As we explain in details in \ref{app1}, in this limit the two lowest orders of the expansion in $\xi$ take the form
\begin{eqnarray}
  \label{lhy2d}
  e_{LHY}^{2{d}}(\xi) \simeq  c_2 \xi^2 + c_3 \xi^3\, ,
\end{eqnarray}
where $c_2 \simeq  0.1974$, $c_3 \simeq 108$. Comparing with the numerical result, we find that this approximation works well as long as $\xi \lesssim 0.002$. The first term of the expansion in Eq.~(\ref{lhy2d}) is proportional to the square of the density and provides a correction to the mean field energy of the BEC.  This correction originates from the effect of the confinement on the two-body scattering amplitude and could also be derived from the two-body problem employing the Born
expansion, as observed also in Refs.~\cite{zin2018droplets,buechler2018crossover}.  The second term proportional to $n^3$ can be interpreted as an emergent three-body
term in the energy functional stemming from quantum fluctuations. This effect is distinct from three-body forces induced by confinement or internal structure, which have been studied e.g. in~\cite{Johnson2009,PetrovPRL2014}. In our case, this term turns out to be repulsive ($c_3>0$), providing a possible stabilization mechanism for the gas close to the instability, and indicating that dipolar droplets may exist in a quasi-2D system. We note that similar effective term can be calculated using perturbation theory on a weakly interacting few-body system~\cite{Petrov2021}.

The expansion in Eq.~(\ref{lhy2d}) has a vastly different structure than the beyond-mean-field term in quasi-2D Bose-Bose mixtures~\cite{zin2018droplets} or in a single
quasi-2D BEC~\cite{buechler2018crossover} with contact interactions.  The origin of this discrepancy can be traced back to the fact that the Fourier transform of the
potential $v(\q)$ near the critical point is linear in $q_\perp$ (for $q_z =0$).  This is the reason why the logarithmic terms that are usually present in two dimensions do not appear here.

Let us now briefly discuss the typical experimental conditions and the parameters needed to reach the 2D regime in the LHY term. Most experiments in this field are performed using dysprosium and erbium~\cite{Boettcher2020} at gas densities $n\sim10^{14}$cm$^{-3}$ with scattering length being of the order of $100\,a_0$. Assuming the box size of $1\mu$m, we then obtain $\xi\approx 1$ for a typical experiment. In order to study the 2D regime one would need to increase the confinement strength by more than an order of magnitude, which could be realized using subwavelength traps~\cite{Wang2018} or a potential shaped by a digital micromirror device~\cite{Tajik19}. Equivalently, one could decrease the gas density to about $10^{11}$cm$^{-3}$, which would on the other hand require much lower temperatures and longer operation times.

\subsubsection{Before the critical point }

In the considerations above, we have so far discussed the LHY term calculated at the critical point $\theta = \pi/2$ and $\epsilon_{dd} = 1$. Below we show that LHY energy in the region of parameters $\delta \theta \ll 1$
and $\delta \epsilon_{dd} \ll 1 $ does not change much. 
We note that, strictly speaking, the LHY energy cannot be calculated 
{\it after} crossing the critical point. Therefore we calculate numerically the 
LHY energy {\it before} and at the critical point in three cases: $\delta \theta = 0$, $\epsilon_{dd} = 1$, $\delta \theta = 0$,   $\epsilon_{dd} =1 - 0.01$, and $\delta \theta = 0.1$, $\epsilon_{dd} = 1$. The results are plotted in the lower panel of Figure~\ref{fig2}. We notice that all the studied cases show the same trend and differ by at most few percent. We have verified that the correction changes continuously as we depart from the $\delta \theta = 0$, $\epsilon_{dd} = 1$ point.
Therefore, in the calculations of the droplet state below we take the value of LHY energy
calculated at the critical point $\epsilon_{dd} =1$ and $\delta \theta =0$.

Finally, we note that for the chosen orientation of the dipoles causing the phonon instability to occur we were able to provide a universal result in the sense that the
expansion coefficients in Eq.~(\ref{lhy2d}) do not depend on the box width. This is related to the fact that for dipoles oriented in plane the condensate depletion
converges, while for perpendicular orientation, where the roton instability occurs, the condensate depletion calculated within Bogoliubov theory diverges and the system
becomes nonuniversal~\cite{Boudjemaa2013,Fedorov2014,jachymski2018nonuniversal}.

\section{The droplet state}
\label{sec:droplet}

So far we have shown that the quasi-2D dipolar gas can in principle support droplet solutions, as the LHY correction provides the mechanism for stabilization. 
In this section, we find the droplet density in the limit of large atom number and numerically investigate the density profile of the droplets in the finite system case.

From Eqs.~(\ref{energia2}) and (\ref{LDA}) we obtain the energy of the considered system where
LHY energy correction is incorporated using local density approximation
\begin{eqnarray} \nonumber
  E[\psi_\perp(\x)]\!=\!&&\int\!\! d\x_\perp \, \frac{\hbar^2}{2m} |\nabla_\perp \psi_\perp|^2 \!+ \frac{1}{2} \! \int \!\!d\x_\perp d\x_\perp'\,  v_{2{d}}(\x_\perp-\x_\perp')|\psi_\perp(\x_\perp)|^2 |\psi_\perp(\x_\perp')|^2 
\\  \label{dlugie}
  &&+\int d \x_\perp \, \frac{\epsilon_0}{L^2} e_{LHY}^{2{d}}\!\left[\frac{2}{\pi} aL |\psi_\perp(\x_\perp)|^2 \right],
\end{eqnarray}
where by $\x_\perp = x {\bf e}_x + y {\bf e}_y$ we denote the two-dimensional position vector.
We note that the coefficient multiplying the square of the wave function in the LHY part is
introduced because we have calculated the correction as a function of $\xi$ instead of the density alone (we remind that here $a$ is the scattering length). 

Now we focus on the $\delta \theta =0$ case.
To calculate the energy we need to have $e_{LHY}^{2{d}}$ as the function of $\epsilon_{dd}$.
In the previous Section we have presented calculation for $\epsilon_{dd} = 1$ and 
$\epsilon_{dd} = 1 -0.01$. We found that $e_{LHY}^{2{d}}$ weakly depends on this change of
$\epsilon_{dd}$. The calculation of LHY in the uniform system case 
are not possible within standard Bogoliubov above the 
critical point i.e. for $\epsilon_{dd} > 1$, due to the imaginary frequencies of low modes.
Still, it is shown, using more sophisticated techniques, that generally $e_{LHY}^{2{d}}$
changes very weakly after crossing the critical point \cite{Ota2020,Hu2020}.
Therefore in what follows in the vicinity of $\epsilon_{dd} =1 $ we approximate 
 $e_{LHY}^{2{d}}$ by its value at $\epsilon_{dd} =1 $.

We now mention an interesting property. This comes from the analysis
of the interaction energy which in the Fourier space takes the form
\begin{eqnarray*}
  \int \!\!d\x_\perp d\x_\perp'\,  v_{2d}(\x_\perp-\x_\perp')
  |\psi_\perp(\x_\perp)|^2 |\psi_\perp(\x_\perp')|^2  
  = \frac{g}{(2\pi)^2} \int\!\!d \K_\perp \, v_{2d}(\K_\perp) [n_{2d}(\K_\perp)]^2,
\end{eqnarray*}
where on the right-hand side $n_{2\mathrm{d}}$ and $v_{2\mathrm{d}} $ are the Fourier transforms of the two-dimensional density, i.e., $n_{2d}(\K_\perp) =\int d \x_\perp
\, e^{-i \K_\perp \x_\perp}|\psi_\perp(\x_\perp)|^2$ and interaction potential
\begin{eqnarray*}
  v_{2d}(\K_\perp) = \int d \x_\perp \, e^{-i \K_\perp \x_\perp} \frac{v_{2d}(\x_\perp)}{g}.
\end{eqnarray*}
The latter can be calculated analytically and for $\delta \theta =0$ reads
\begin{eqnarray}\label{v2dFT}
  v_{2d}(\K_\perp) \!=\! \frac{1}{L} \left\{ 1 \! +\! \epsilon_{dd} \left[  \frac{3 k_x^2}{k_\perp^3L}
    \left(  e^{ - k_\perp L} \!+\! k_\perp L  \!-\!1  \right)  \!-\!1 \right]  \right\}. \quad
\end{eqnarray}
The form of the potential from Eq.~(\ref{v2dFT}) suggests it can be split into two parts. The first one is $v_{2{d},loc}(\K_\perp) = \frac{1 - \epsilon_{dd}}{L}$, whose inverse Fourier transform yields a contact (local) potential $v_{2{d},loc}(\x_\perp) = \frac{1 - \epsilon_{dd}}{L} g \delta(\x_\perp)$.
The second part is the nonlocal potential
\begin{equation}\label{vnon}
v_{2{d},non}(\K_\perp) =  \frac{\epsilon_{dd}}{L}    \frac{3 k_x^2}{k_\perp^3L}
\left(  e^{ - k_\perp L} + k_\perp L  -1  \right).
\end{equation} 
We notice that this potential goes to zero for $k_\perp \rightarrow 0$, and so we expect it to be less important as the size of the droplets grows.

Having the above we now focus on the quasi-2D limit.
Here we make use of  expansion given by Eq.~(\ref{lhy2d}).
Firstly, we calculate the equilibrium density $|\psi_\perp|^2 =n^{eq}$ of a large droplet.
Here we make a simple model of the droplet as having constant density and volume
 in the two-dimensional plane denoted as $V_\perp$. We assume that the droplet finishes sharply and neglect the energies of the boundaries.
In such a case, the energy of the droplet is 
\begin{equation} 
   E =  \frac{g}{2L} (1-\epsilon_{dd}) (n^{eq})^2 V_\perp   +
  \frac{\epsilon_0}{L^2} \left( c_2 \left(\frac{2}{\pi} aL  n^{eq }\right)^2 +  
  c_3 \left(\frac{2}{\pi} aL  n^{eq}\right)^3 \right) V_\perp.
   \label{Ee}
\end{equation}
Note that above we use the approximation described above and neglect $v_{2{d},non}$.
Substituting here the normalization condition, i.e., $N = n^{eq} V_\perp$,
and requiring $(\partial E/\partial V_\perp)_N = 0$, we find the equilibrium density 
of the droplet
\begin{eqnarray*}
	\label{eq:eq}
 n^{eq} =   \frac{1}{ a^2}   \frac{\pi^2 \epsilon}{16 c_3}.
\end{eqnarray*}
Here $ \epsilon = \epsilon_{dd} - 1 - \frac{4}{\pi} \frac{a}{L} c_2$ is the stability parameter, and the droplet is formed when $\epsilon>0$. We notice that the critical point is shifted from the usual condition $\epsilon_{dd} - 1>0$ by the correction to the mean field coupling strength due to confinement. 


With the equilibrium density at hand, we rewrite the energy functional in the convenient dimensionless units, i.e.,  the unit of length $d = \frac{\sqrt{2 c_3}}{\pi^{3/2} \epsilon
} \sqrt{aL}$, the unit of energy $E_0 = \frac{g}{L} \epsilon ( n^{eq})^2 d^2$, and we set $\psi_\perp/\sqrt{n^{eq}} \rightarrow \psi_\perp $.
Such a transformation results in
\begin{equation}\label{gpe2d}
  E \!=\! \int \!\!d\x_\perp  \bigg[|\nabla_\perp \psi_\perp(\x_\perp) |^2 \!-\! \frac{1}{2} |\psi_\perp (\x_\perp)|^4
  \!+\! \frac{1}{4} |\psi_\perp(\x_\perp)|^6 \bigg] \!+\! \delta E,
\end{equation}
where the first integral contains the local terms and $\delta E$ is the contribution to the total energy from the nonlocal part of the dipole-dipole term
\begin{eqnarray*}
  \delta E = \frac{L}{8\pi^2 \epsilon} \int d \K_\perp \,  v_{2\mathrm{d},non}\left(\frac{\K_\perp}{d} \right) |n_{2d}(\K_\perp) |^2.
\end{eqnarray*}
We note that here $n_{2d}(\K_\perp)$ is the Fourier transform of the density (defined as before but now in the new units), and $v_{2\mathrm{d},non}$ is given by Eq.~(\ref{vnon}).
The physical number of atoms in the droplet is 
\begin{eqnarray*} 
  N_{at} = n^{eq} d^2 N  =  \frac{L}{a} \frac{1}{8\pi \epsilon} N,
\end{eqnarray*}
where $N = \int d \x_\perp \, |\psi_\perp(\x_\perp)|^2$.

We have minimized the functional from Eq.~(\ref{gpe2d}) numerically and found the density profiles of the droplet for different total number of atoms.
In the numerical calculations we take $a = 10\ \mathrm{nm}$, the length $L = 1\ \mu\mathrm{m}$, and we set $\xi^{eq} = 0.001$. 
These numbers lead to $\epsilon = \xi^{eq} \frac{a}{L} \frac{8c_3}{\pi} \simeq 0.0027 $
and $ n^{eq} d^2 = \frac{L}{a} \frac{1}{8\pi \epsilon} \simeq 1400 $.
In order to find the density that minimizes the function, we calculate the functional derivative of
$E -  \mu N$ with respect to $\psi_\perp(x,y)$; here, we impose an additional constraint on the total norm so a Lagrange multiplier $\mu$ appears. This
procedure leads to a Gross-Pitaevskii-type equation for $\psi_\perp$. In the first step of our numerical approach, we drop from the functional the term $\delta E$, and
then we solve it by the imaginary-time method. This leads to a profile $\psi_\perp$ which is spatially symmetric. This solution is then treated as the initial point for the full problem including now the term $\delta E$.

\begin{figure}[tb]
  \centering \includegraphics[width=0.6\columnwidth]{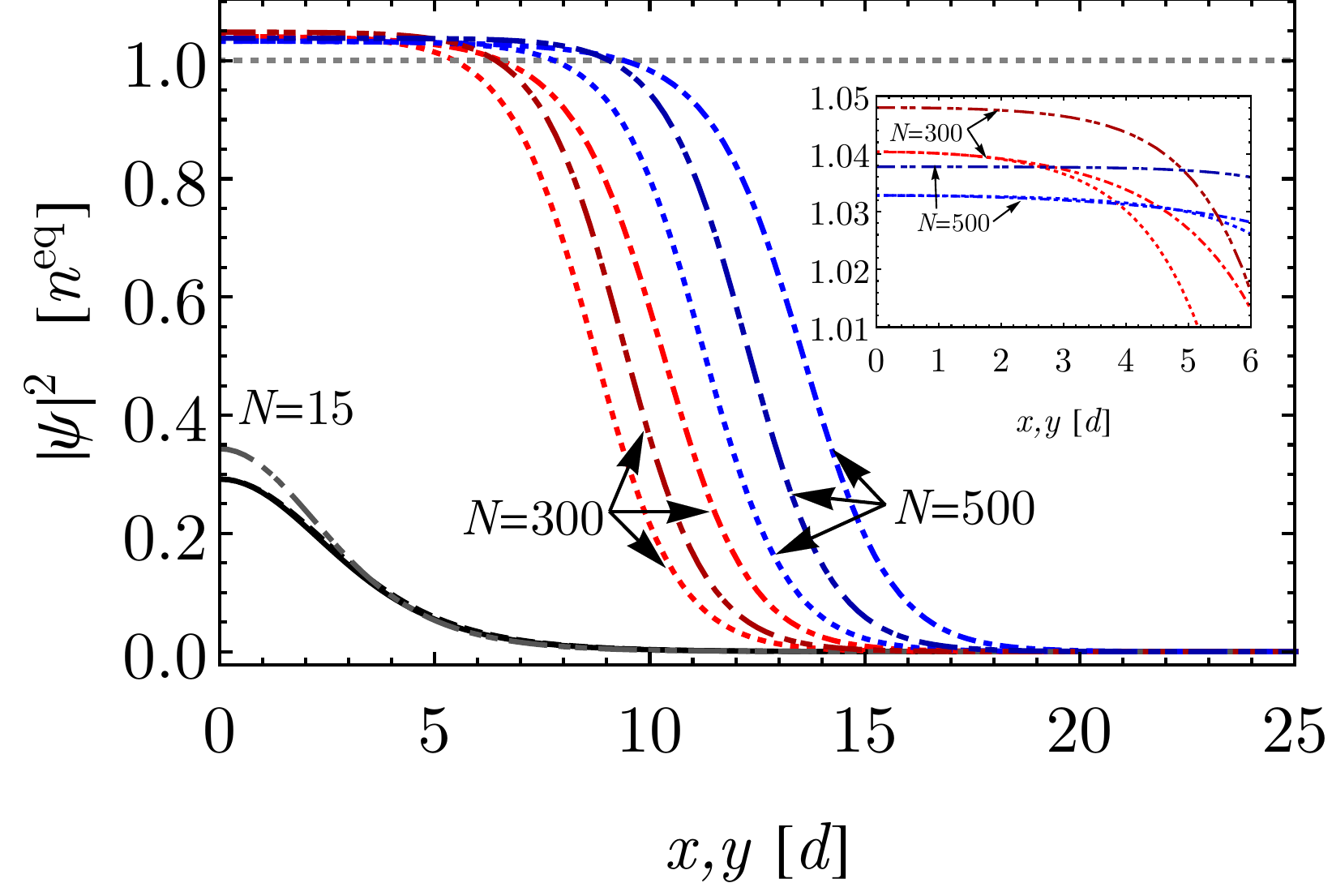}
  \caption{The cuts through the two-dimensional density of the droplet along the $x$ and $y$ directions taken through the centre of the cloud $|\psi_\perp(x,0)|^2$ and $|\psi_\perp(0,y)|^2$. 
    The black solid line is calculated for $N=15$; here the two cuts are indistinguishable and no anisotropy can be seen.
    The red dotted (red dotted-dashed) and blue dotted (blue dotted-dashed) lines are for $N=300$ and $500$, respectively, and the cuts along $y$ ($x$) direction; 
    the norm $N$ is also indicated by arrows.
    In each case, the double-dotted-dashed line between the two cuts shows the result of the calculation without the anisotropic contribution $\delta E$ which gives a symmetric profile. 
    The inset (the units on axes are the same as in the main panel) shows the zoom in of the cloud center. 
    The symmetric solution (without $\delta E$) always overestimates the central density. With increasing the norm $N$, the full solution approaches the homogeneous limit.
    In the main panel, the horizontal, dotted gray line indicates the homogeneous limit~$|\psi_\perp|=n^{eq}$.
}
\label{fig_droplet}
\end{figure}

Fig.~\ref{fig_droplet} displays the numerically calculated density profiles of
the droplets for three values of the two-dimensional norm $N = 15$, 300 and 500. This corresponds to the number of atoms: $N_{at} = 2.1 \times 10^4$, $4.2 \times 10^5$ and $7.0\times 10^{5}$, respectively.
We note that for these parameters the three-dimensional
density of atoms is on the order of $10^{11} \mathrm{cm}^{-3}$, and so the three-body losses not included in the theory should still be moderate.

As can be seen in Fig.~\ref{fig_droplet}, due to the effect of $\delta E$, the droplet shape that we obtain is not cylindrically symmetric. The observed anisotropy, however, is not
prominent. For larger atom numbers, for instance $N = 300$ or 500, the density in the middle is almost constant and close to the analytically predicted equilibrium value
$n^\mathrm{eq}$ (see also the inset of Fig.~\ref{fig_droplet}). Finally, we see that with increasing the number of atoms, the droplet grows in its volume by keeping almost constant
central density, but attaching the atoms mainly to its surface. This effect indicates that the system has liquid properties.



Now we shortly discuss the validity of the assumptions made in Sections~\ref{sec:description} and~\ref{sec:LHY}.  In Sec.~\ref{sec:description}, we assumed that the length on which
the density changes in the $x$ and $y$ directions is much larger than $L$. From our analysis, we find that this characteristic lengthscale is given by $d$, and the required
condition is $d \gg L$.  In the case of quasi-2D limit analyzed in this Section, we have $\frac{L}{d} = \sigma^{eq} \sqrt{\frac{a}{L}} 4 \sqrt{2 \pi c_3 }$, and using the values of
the parameters taken from numerical simulations, we obtain $L/d \simeq 0.01$, which confirms the separation of lengthscales.

Additionally, in Sec.~\ref{sec:LHY} we assumed the validity of the LDA.  When we calculated the LHY energy in the uniform case, we integrated and summed over the wave vectors $\K$ with a
characteristic cutoff $k_c$. In the quasi-2D geometry, $k_c$ is of the order of a few $1/L$, but the coefficients $c_2$ and $c_3$ have their values approximately the same as for
$k_c = \infty$.  The LDA is justified, when $k_c$ is much larger than the inverse of the length on which the density changes in the $x$-$y$ plane, which we found to be equal to
$d$. Therefore, in the quasi-2D limit, we arrive at $d \gg L$, which is the same condition as obtained above where we analyzed the assumptions stated in Sec.~\ref{sec:description}.


Finally, we discuss the droplet formation in the case $\delta \theta \neq 0$ and  $\delta \theta \ll 1$.
This problem can be found by minimization of the energy functional given by Eq.~(\ref{dlugie}).
In the above we did that for $\delta \theta =0$.
Still for $\delta \theta \ll 1$ the potential $v_{2d}(\x_\perp)$ shall not change much (as it is a continuous 
function of $\delta \theta$) with respect to $\delta \theta =0$ case.
In addition we have shown that the change of LHY energy term for $\delta \theta \ll 1$ is negligible. 
The droplet state comes from the interplay between interaction and LHY energy. 
As   both of these does not change much for $\delta \theta \ll 1$ thus the properties of the
droplet state shall also be close to those analysed above ($\delta \theta =0$ case).


\section{Conclusions}
\label{sec:conclusion}

We analysed the beyond-mean-field behaviour of dipolar Bose gas subject to quasi-two-dimensional confinement. Under several simplifying assumptions, we determined analytically the Lee-Huang-Yang correction to the mean field energy.  We showed that close to
the phonon instability the correction can become decisive for the properties of the system, preventing it from collapse. The ground state of the system in this case is a
finite size self-bound droplet similar to the three-dimensional case. Crucially, we found that for moderate confinement strength the magnitude of the correction is close to its free-space limit, validating the use of the local density approximation.


\ack
P.Z. and Z.I. acknowledge the support from the Polish National Science Centre Grant No. 2015/17/B/ST2/00592.  
M.P. acknowledges support from grant No. 2017/25/B/ST2/01943.

\appendix

\section{Evaluation of the Fourier transform of the dipole-dipole interaction potential}
\label{app0}

In the case of the 3D dipole-dipole interaction the Fourier transform of
$v_{dip}(\x) = g \epsilon_{dd} \frac{3 }{4\pi} \frac{1 - 3 ({\bf e}_\x \cdot {\bf e})^2}{r^3}$
(where ${\bf e}_\x = \x/|\x|$ and ${\bf e}$ denotes the direction of dipole polarization)
is evaluated by
taking the integral in a finite space region between two spheres of radius $r_0$ and $R_0$. After performing the integral, the limit $r_0 \rightarrow 0$ and $R_0 \rightarrow \infty$ is taken arriving at 
$ v_{dip,3d}(\K) =  \epsilon_{dd} \left( 3 (\K \cdot {\bf e})^2/k^2-1 \right) $.
Here we need to perform the integral
\begin{equation}\label{ap11}
  v_{dip}(\K) = \int_{-L/2}^{L/2} \mbox{d} z \int \mbox{d} \x_\perp \, e^{-i\K\x} \frac{v_{dip}(\x)}{g}
\end{equation}
where $k_z = \frac{2\pi}{L} n_z$ is quantized.
As we are restricted to the quasi-2d space, we do not need to use the cutoff at large distances
and we only take a cutoff at small distances $r_0$.
This can be done by taking
\begin{equation}\label{ap12}
  v_{dip}(\x) = \frac{1}{(2\pi)^3}\int \mbox{d} \K' \, e^{i \K' \x} v_{dip,3d}(\K') h(k') .
\end{equation}
where $h(k')$ is a function being equal to unity for small $k'$ and then going to zero for 
$k'> k_0$ where $k_0 \simeq 1/r_0$. Finally we take the $k_0 \rightarrow \infty$ limit.
Inserting Eq.~(\ref{ap12}) into Eq.~(\ref{ap11}) leads to
\begin{eqnarray*}
  v_{dip}(\K) = \frac{1}{(2\pi)^3} \int_{-L/2}^{L/2} \mbox{d} z \int \mbox{d} \x_\perp \, 
  e^{-i\K\x} 
  \int \mbox{d} \K' \, e^{i \K'\x} v_{dip,3d}(\K') h(k').
\end{eqnarray*}
Performing the above analytical integrals gives
\begin{eqnarray*}
  v_{dip}(\K) = \frac{1}{2\pi} \int_{-L/2}^{L/2} \mbox{d} z \int \mbox{d} k_z' \, e^{-i(k_z-k_z')z}  v_{dip,3d}(k_x,k_y,k_z') 
  h \left(\sqrt{k_x^2+k_y^2+{k_z'}^2} \right).
\end{eqnarray*}
In the considered geometry $v_{dip,3d}(\K) = \epsilon_{dd}(3(k_x\sin \theta + k_z \cos \theta)^2/k^2-1) $. Inserting $v_{dip}$ into the above we obtain
\begin{eqnarray*}
&& \fl  v_{dip}(\K) = \frac{1}{2\pi} \int_{-L/2}^{L/2}\mbox{d} z \int \mbox{d} k_z' \, e^{-i(k_z-k_z')z} 
 \epsilon_{dd} \left(  \frac{3 (k_x\sin \theta + {k_z'} \cos \theta)^2}{k_x^2+k_y^2 + {k_z'}^2} -1 \right)
  h \left(\sqrt{k_x^2+k_y^2+{k_z'}^2} \right).
  \\
  \\
  &&  \fl
  = \frac{1}{2\pi} \int_{-L/2}^{L/2} \mbox{d} z \int \mbox{d} k_z' \, e^{-i(k_z-k_z')z} 
  \epsilon_{dd} \left(  3 \cos^2 \theta - 1
  +  3
  \frac{k_x^2\sin^2 \theta  - (k_x^2 + k_y^2) \cos^2 \theta+ 2 k_x k_z' \sin\theta \cos \theta}{k_x^2+k_y^2 + {k_z'}^2}
  \right)
\\
\\
&& \fl \times   h \left(\sqrt{k_x^2+k_y^2+{k_z'}^2} \right).
\end{eqnarray*}
In the above we deal with three different kinds of integrals. The first one
\begin{eqnarray*}
  \frac{1}{2\pi} \int_{-L/2}^{L/2} \mbox{d} z \int \mbox{d} k_z' \, e^{-i(k_z-k_z')z} 
  (3 \cos^2 \theta - 1)h \left(\sqrt{k_x^2+k_y^2+{k_z'}^2} \right)
\end{eqnarray*}
simply equals to $ 3 \cos^2 \theta - 1 $ after taking the limit $k_0 \rightarrow \infty$. To see it clearly one can take $h(k') = \exp(-{k'}^2/k_0^2)$, perform the integrals analytically and at the end take the limit.
The second integral is
\begin{eqnarray*}
  \frac{1}{2\pi} \int_{-L/2}^{L/2} \mbox{d} z \int \mbox{d} k_z' \, e^{-i(k_z-k_z')z} 
  3
  \frac{k_x^2\sin^2 \theta - (k_x^2 + k_y^2) \cos^2 \theta}{k_x^2+k_y^2 + {k_z'}^2}
  h\left(\sqrt{k_x^2+k_y^2+{k_z'}^2} \right).
\end{eqnarray*}
Here the integrand goes as $1/{k_z'}^2$ for large $k_z'$ . Thus we may readily
set $h = 1$ obtaining an analytical result that reads
\begin{eqnarray*}
     3 \frac{k_x^2 \sin^2 \theta - k_\perp^2 \cos^2 \theta}{k^2 } \left( 1- \exp \left( - \frac{k_\perp L}{2}  \right) (-1)^{ k_z L/(2\pi)}   \right) 
\end{eqnarray*}
where $k_\perp^2 = k_x^2+k_y^2$ and $k_z = \frac{2\pi}{L} n_z$.
The third kind of integral reads
\begin{eqnarray*}
  \frac{1}{2\pi} \int_{-L/2}^{L/2} \mbox{d} z \int \mbox{d} k_z' \, e^{-i(k_z-k_z')z} 
  3
  \frac{ k_x k_z' \sin 2\theta}{k_x^2+k_y^2 + {k_z'}^2}
  h\left(\sqrt{k_x^2+k_y^2+{k_z'}^2} \right).
\end{eqnarray*}
Setting $h=1$ we obtain an analytic result that reads
\begin{eqnarray*}
3\frac{ k_x k_z \sin 2 \theta}{ k^2} 
\left( 1 - \exp \left( - \frac{k_\perp L}{2}  \right)   (-1)^{ k_z L/(2\pi)} \right)
\end{eqnarray*} 
Adding the above  together we arrive at
\begin{eqnarray*}
\fl
v_{dip}(\K) =  \epsilon_{dd} \left(   3 \cos^2 \theta -1 
+ 3  \frac{k_x^2 \sin^2 \theta + k_x k_z \sin 2 \theta  - k_\perp^2 \cos^2 \theta}{k^2}  \left( 1- \exp \left( - \frac{k_\perp L}{2}  \right) (-1)^{ k_z L/(2\pi)}   \right) \right). 
\end{eqnarray*}

\section{Derivation of the formula for $e_{LHY}^{2d} $ }  \label{appR}

Here we start from analyzing the system of atoms interacting via 
potential $\tilde v(\x)$. We choose as an effective potential a function of the form
\begin{equation}\label{tildev}
  \tilde v(\x) =  v_c(\x) +  g f(|\x|) \frac{3 \epsilon_{dd}}{4 \pi r^3} \left( 1 - 3 \cos^2 \theta \right) = v_c(\x) + v_{dip}(\x).
\end{equation}
In the above equation $v_c(\x)$ is a non-negative central and local potential of width $\sigma$.
In Eq.~(\ref{tildev}) we additionally introduced a function $f(r)$ which is equal to unity for 
$r_0 < r < R_0$ and goes to zero otherwise, where $r_0$ and $R_0$ are the small and large distance regularization
used in \ref{app0}.

Applying Bogoliubov method to the homogeneous system in the same geometry as in the main part of the paper we obtain the ground state energy density equal to
\begin{equation}\label{e0}
e_0 = \frac{1}{2} n^2 g \tilde v(\K=0) + \frac{1}{2(2\pi)^2 L} \sum_{k_z}  \int \mbox{d} \K_\perp \,  \left( \varepsilon(\K) - E_k - n g \tilde v(\K) \right) 
\end{equation}
where $\varepsilon(\K)  = \sqrt{ E_k(E_k + 2 n g \tilde v(\K) ) }$
and $\tilde v (\K)  =  \int_{-L/2}^{L/2} \mbox{d} z  \int \mbox{d} \x_\perp \, e^{- i \K \x}  \tilde v(\x)/g$. Additionally we define $\tilde v_c(\K)$ and $\tilde v_d(\K)$ in the same way as $\tilde v(\K)$.
In the above the ground state energy is given as a function of $\tilde v(\K)$.
Now we have to relate $\tilde v(\K)$ to the universal quantities like the scattering length.
In order to do it we use Born expansion of the potential up to the second order
\begin{equation}\label{g}
g = g \tilde v_{3d}(\K=0) - \frac{1}{(2\pi)^3} \int \mbox{d} \K \, \frac{g^2 \tilde v_{3d}^2(\K)}{2 E_k} + \ldots
\end{equation}
where $\tilde v_{3d}(\K) = \int \mbox{d} \x \, e^{- i \K \x}  \tilde v(\x)/g $.
As before we additionally defined $\tilde v_{c,3d}(\K)$ and $\tilde v_{d,3d}(\K)$.
We fist notice that $\tilde v_{d,3d}(\K=0) = 0$ which results in  $\tilde v_{3d}(\K=0) = \tilde v_{c,3d}(\K=0) $. 
Now we assume that the potential $v_c(\x)$ is negligible for $r > L/2$.
This implies that $\tilde v_c(\K) = \tilde v_{c,3d}(\K)$.
We thus obtain
\begin{equation} \label{rel}
\tilde v(\K=0) = \tilde v_{3d}(\K=0) +  \tilde v_d(\K=0).
\end{equation}
Combining Eqs.~(\ref{e0}), (\ref{g}) and (\ref{rel}) leads to
\begin{equation}\label{e02}
\fl
e_0 = \frac{1}{2} n^2 g \left( 1 + \tilde v_d(\K=0)  \right)
 + \frac{n^2}{2(2\pi)^3} \int \mbox{d} \K \, \frac{g^2 \tilde v_{3d}^2(\K)}{2 E_k} 
+\frac{1}{2(2\pi)^2 L} \sum_{k_z}  \int \mbox{d} \K_\perp \,  \left( \varepsilon(\K) - E_k - n g\tilde v(\K) \right) .
\end{equation}
In the above we identify the LHY energy density which reads
\begin{eqnarray*}
e_{\rm LHY}= \frac{n^2}{2(2\pi)^3} \int \mbox{d} \K \, \frac{g^2 \tilde v_{3d}^2(\K)}{2 E_k} 
-\frac{1}{2(2\pi)^2 L} \sum_{k_z}  \int \mbox{d} \K_\perp \,  \frac{n^2 g^2 \tilde v^2(\K)}{\varepsilon(\K) + E_k + n g \tilde v(\K)}
\end{eqnarray*}
where we additionally used the relation $ - \frac{n^2 g^2 \tilde v^2(\K)}{\varepsilon(\K) + E_k + n g\tilde v(\K)} = \varepsilon(\K) - E_k - n g\tilde v(\K) $.

We now assume that the width of the $v_c(\x)$ potential $\sigma$ is much larger than the scattering length $a_c$ of the $v_c$ potential (which is positive as $v_c(\x) \geq 0$),
and additionally much larger than $a$, i.e., $\sigma \gg a_c$, $\sigma \gg a$.
These assumptions imply that for $|\K| \ll \frac{1}{\sigma}$ we have 
$\tilde v(\K) \simeq  v(\K) $ which in fact justifies the use of pseudopotential $v(\x)$ given by
Eq.~(\ref{v}). 
In the above we still have $r_0$ and $R_0$ present which make the above quantities finite.
We want to take the limit $r_0 \rightarrow 0$ and $R_0 \rightarrow \infty$.
In such a case we need to define how we take the limit.
We do it taking 
\begin{eqnarray*}
\fl
e_{\rm LHY}= -   
\frac{1}{2(2\pi)^2 L} \sum_{k_z} 
 \int \mbox{d} \K_\perp \,
\left(   \frac{n^2 g^2 \tilde v^2(\K)}{\varepsilon(\K) + E_k + n g \tilde v(\K)}
-   
\frac{L}{2\pi}    \int_{k_z- \Delta k_z/2}^{k_z + \Delta k_z/2} \mbox{d} k_z' \,
\frac{g^2 \tilde v_{3d}^2(k_x,k_y,k_z')}{\frac{\hbar^2}{m} (k_x^2 + k_y^2 + {k_z'}^2)}
\right)
\end{eqnarray*}
We now take the limit 
which gives $\tilde v_{d,3d}(\K) = v_{3d}(\K)$ where $v_{3d}(\K) = 1 + \epsilon_{dd} \left( \frac{k_x^2}{k^2} -1 \right) $. 
After doing that we equate the above to $ \frac{\epsilon_0}{L^3} e_{LHY}^{2d}(\xi) $, substitute into the above the definition of $\xi$ and $\q$.
As a result we  arrive at Eq.~(\ref{lhydef}) from the main text.

\section{Evaluation of the quasi-2D limit}
\label{app1}
In this section we evaluate analytically the quasi-2D limit of the LHY correction.
We rewrite Eq.~(\ref{lhydef}) as:
\begin{eqnarray*}
  e_{LHY}^{2d}  = 
  - \frac{\xi^2}{2} \sum_{q_z}\!\! \int\!\! \mbox{d} \q_\perp   \frac{v^2(\q)}{ \epsilon_\q + q^2 + \xi v(\q)} \!-\!  \int\!\! \mbox{d}\, \q \frac{ v_{3d}^2(\q)}{2q^2} =  -\frac{\xi^2}{2} \left( g(\xi)  -2 c_2 \right)
\end{eqnarray*}
where
\begin{eqnarray*}
  && g(\xi) = \sum_{q_z } \int \mbox{d} \q_\perp 
  \left( \frac{v^2(\q)}{ \epsilon_\q + q^2 + \xi v(\q)} - \frac{v^2(\q)}{ 2q^2}  \right)
  \\
  && -2 c_2 = \sum_{q_z } \int \mbox{d} \q_\perp   \frac{v^2(\q)}{ 2q^2}  -  \int \mbox{d} \q \frac{v_{3d}^2(\q)}{2q^2}. 
\end{eqnarray*}
Expanding function $g(\xi)$ around $\xi=0$, we obtain
\begin{eqnarray*}
  g(\xi)   \simeq - \xi   \sum_{q_z} \int \mbox{d} \q_\perp \, \frac{v^3(\q)}{2q^4}
  = - 2 c_3 \xi
\end{eqnarray*}
In the above we notice that the two constants $c_2$ and $c_3$ are given by convergent sums and can be evaluated to obtain
\begin{eqnarray*}
&& c_2 =   \frac{27}{256} \pi ( 3 -  2 \zeta(3) ) \simeq 0.1974
\\
&& c_{3} = \frac{135 \pi^3}{8192} \left(  167 + 6 \pi^2 - 12 \zeta(3) \right) \simeq 108.
\end{eqnarray*}

\providecommand{\newblock}{}


\end{document}